\documentclass[aps,pre,preprint,showpacs,preprintnumbers,amsmath,amssymb]
{revtex4}
\usepackage{graphicx}
\usepackage{bm}

\def\be{\begin{equation}}
\def\ba{\begin{eqnarray}}
\def\a{\alpha}
\def\b{\beta}

\def\d{\delta}
\def\D{\Delta}
\def\e{\epsilon}

\def\th{\theta}

\def\p{\pi}
\def\r{\rho}

\def\S{\Sigma}
\def\t{\tau}

\def\c{\chi}

\def\o{\omega}
\def\O{\Omega}
\def\i{\int}

\def\bx{{\mathbf x}}
\def\bX{{\mathbf X}}
\def\by{{\mathbf y}}

\def\bD{{\mathbf D}}
\def\bK{{\mathbf K}}

\def\ee#1{\label{#1}\end{equation}}
\def\ea#1{\label{#1}\end{eqnarray}}

\begin{document}
\title{Rate Description of Fokker-Planck Processes
with Time Dependent Parameters }
\author{Peter Talkner}
\affiliation{Universit\"at Augsburg, Institut f\"ur Physik, 
Universit\"atsstrasse 1, D-86135 Augsburg, Germany}
\author{Jerzy \L uczka}
\affiliation{Institute of Physics,  University of Silesia, 40-007
Katowice, Poland}
\date{\today}
\begin{abstract}
The reduction of a continuous Markov process with multiple metastable
states to a discrete rate process is investigated in the presence of
slow 
time dependent parameters such as periodic external forces or
slowly fluctuating barrier heights. 
A quantitative criterion is provided
under which condition a kinetic description with time dependent frozen 
rates applies. Finally it is shown how the long time behavior of the
underlying continuous process can be retrieved from the knowledge of
the discrete process by means of an appropriate random decoration of
the discrete states. As a particular example of the presented theory an 
over-damped bistable Brownian oscillator with periodic driving is
discussed. 
\end{abstract}

\maketitle

\section{Introduction}\label{I}
The separation of time scales is a frequently met dynamical feature of
physical,
chemical and biological
processes. In
practically all cases the parameters characterizing a particular system
are only constant on a limited time scale. On longer time scales they
may change their values because the surrounding environment slowly
changes its state. We just note that even the most fundamental ``constants''
of nature like e.g. the fine structure constant possibly undergo a very slow
change in time \cite{D}. On the other hand,
most natural processes also display very fast changes which usually will be
disregarded in a phenomenological description. They only will show up in a
microscopic description in which the effective degrees of freedom of the
phenomenological model are
represented as functions of microscopic degrees of freedom.
The time evolution of
these microscopic degrees of freedom
typically takes place on much faster time scales than those of the
macroscopic observables whereas, as already mentioned, global aspects of the 
environment may
undergo temporal variations that are much slower than the dynamics of the
considered macroscopic system.

It is this separation of time scales that allows
one to describe a system in terms of a comparatively simple model that
is 
closed in the sense that only a few macroscopic observables are
subject to the dynamics.
On the slow time scale 
the key technical notion
here is that of a
{\it constraint equilibrium} that is reached by the fast part of a system
before the slow constituents change their values on the slower time scale.
This constraint equilibrium then immediately adapts to 
the changes of the slow system.

This approach has been applied
in many branches of physics and
other sciences and is vital for various methods like
adiabatic elimination, averaging,
synergetics, subdynamics, chemical kinetics and hierarchical computational
schemes \cite{ae}. In the present paper we consider Markovian
processes the dynamics of
which can be subdivided into a fast and a slow component.
In particular, we have in mind the case where the slow motion is caused by
transitions between 
metastable states \cite{htb}. The relaxation toward these states
constitutes the fast part of the dynamics.
Additionally, some
of the system's parameters may undergo slow variations in time that may be
caused by a slow drift, or fluctuations of the parameters, or by an
external driving of the system.
The dynamics of the considered system then has two slow
components: The transitions and the driving. 
About their relation we do not make further assumptions:
The slow intrinsic dynamics can be faster or slower than the
external driving that must only be slow compared to the fast dynamics of the
system. In Ref.~\cite{t99} we have called this type of situation  the
semiadiabatic limit, in contrast to the adiabatic limit in which the driving
is slow compared to the total intrinsic dynamics including that of the
transitions. Clearly, the
adiabatic limit is
contained in the semiadiabatic limit as a special case.

Stochastic resonance is just one example for which the semiadiabatic limit
is of relevance when the period of the signal is large
\cite{j,ghjm}. 
Ratchets, or Brownian motors, provide another class of systems 
which also operate in the semiadiabatic regime \cite{r}.
Atomic force microscopes and optical tweezers often act on time
scales that are slow compared to the vibrational and other intrinsic
time-scales of the molecules that are manipulated by them
\cite{erm}.

The paper is organized as follows. In Sect.~\ref{II}
the reduction of
a Fokker-Planck process with multiple mestastable states to a master
equation is reviewed as it was developed in
Refs.~\cite{Ry,Mo}. This method is generalized to processes with 
slowly time dependent parameters in Sect.~\ref{III}. The time rate of
change of the probability now consists of two contributions. One
is given by the rate as it would result if the parameters of the
process were frozen and the other contribution mainly takes into
account the changes of the geometry of the domains of attraction of
the metastable states. This geometric contribution is proportional to
the time rate of change of the parameters and therefore is negligible
compared to the frozen rates if the time dependent parameters change 
sufficiently
slowly. In contrast to previous investigations,
\cite{Ast} we find that the fastest time dependence
of the parameters for which one may neglect the geometric contributions
is not only determined by the deterministic time scales of the process  
but in general also depends on the strength of the noise. The way how
the noise here enters depends on the particular process.

The present investigation is complementary to previous work that is restricted 
to periodic driving with an intermediate regime of external driving 
frequencies \cite{lrh}. 

In Sect.~\ref{IV} we assume the validity of the master equation with
frozen rates and determine a time dependent 
decoration of the discrete states such that the dynamics of the
continuous variables of the underlying Fokker-Planck process are
recovered on the long time scale. The archetypical example for
stochastic resonance namely that of a
periodically damped bistable Brownian oscillator is discussed in 
Sect.~\ref{V} and the validity of the McNamara-Wiesenfeld model of
stochastic resonance 
\cite{McNW} 
is discussed. 
The paper ends with a summary in Sect.~\ref{VI}.

\section{Multistable systems at small noise}\label{II}
In this section we review the reduction of a Fokker-Planck equation 
describing a {\it time-homogeneous} process with multiple metastable states 
to a master equation that gives the dynamics 
on the time scales of the 
transitions between the metastable states.
For this purpose we consider a dynamical system under the 
influence of weak random
forces and assume that these forces can be modeled by Gaussian white noise. 
Then the stochastic dynamics is described by a Fokker-Planck equation 
\cite{Ri}:
\be
\frac{\partial }{\partial t}\r(\bx,t) = L \r(\bx,t)
\ee{FP}
where $L$ denotes the Fokker-Planck operator:
\be
L = -\sum_i \frac{\partial}{\partial x_i} K_i(\bx) + 
\sum_{i,j} \frac{\partial^2}{\partial x_i
\partial x_j} D_{ij}(\bx)
\ee{FPO}
Here, $\bx = (x_1, x_2, ... x_n)$ denotes a point with coordinates $x_i$ in
the $n$-dimensional state space $\S$; further,
$\bK(\bx) =(K_i(\bx))$ is the drift vector and $\bD(\bx) = (D_{i,j}(\bx))$ 
the diffusion
matrix resulting from the random forces. We further restrict ourselves
to systems with a uniquely defined stationary
probability density $\r_0(\bx)$ with respect to which detailed balance
holds. Consequently, the Fokker-Planck operator satisfies the relation
\cite{Ri}:
\be
L \hat{\r}_0 = \hat{\r}_0 \tilde{L}^+.
\ee{db}
where $L^+$ 
is the backward, or adjoint Fokker-Planck operator:
\be
L^+ = K_i(\bx) \frac{\partial}{\partial x_i} + 
D_{i,j}(\bx) \frac{\partial^2}{\partial x_i \partial x_j}.
\ee{BFP}
The tilde denotes the operation of time reversal, i.e. 
$\tilde{x}_i =\e_i x_i$ with parity $\e_i = \pm 1$ and 
$\hat{\r}_0$ is the multiplication operator with the
probability density, i.e. $\hat{\r}_0 f(\bx) = \r_0(\bx)f(\bx)$ where $f(\bx)$
is an arbitrary 
state space function. 

In the deterministic limit the
diffusion matrix goes to zero and the drift vector approaches the
deterministic vector field $\bK^{(0)}(\bx)$ governing the deterministic motion
of the system:
\be
\dot{\bx}(t) = \bK^{(0)}(\bx(t))
\ee{dem}
In the presence of weak random perturbations $\bK(\bx)$ 
may differ from $\bK^{0}(\bx)$
by small noise induced contributions.

We here are interested in cases where the deterministic system
(\ref{dem}) has a number $m$ of 
coexisting attractors labeled by $\a$. To each attractor
$\a$ there belongs
a domain of attraction ${\cal D}_\a$.
The domains are disjoint and partition the total available state space:
\ba
{\cal D}_\a \bigcap {\cal D}_{\a'} & = & \emptyset \quad \mbox{for} 
\; \a \neq {\a'} \noindent \\
\bigcup_{\a =1}^m {\cal D}_\a & = & \S
\ea{da}
\begin{figure}
\includegraphics[width=10cm]{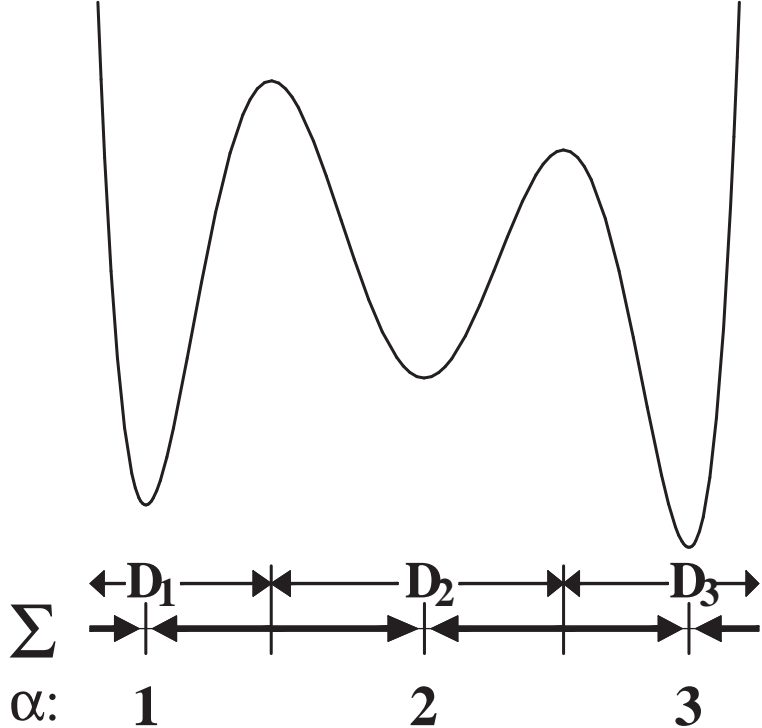}
\caption{The overdamped deterministic motion $\dot{x}(t) = - V'(x(t))$
in the above shown potential induces  a  
partitioning of the state space $\S = \mathbb{R}_1$ into the
domains of attraction $\mathcal{D}_\a$  of the three 
attractors $\a =1,2,3$ at the potential minima. Here the locations of
the local potential maxima coincide with the boundaries between the
domains of attraction.}
\label{f1}
\end{figure}
For weak noise the stationary probability density is almost zero
everywhere except close to the attractors. If the noise vanishes the
stationary density shrinks to Dirac $\d$-functions sitting at the attractors
of the deterministic system. Because no transitions between the different
attractors may occur in the noiseless case, each of the delta functions is a
stationary solution of the Fokker-Planck equation
$\partial/ \partial  x_i K_i^0(\bx) \r_0(\bx) = 0$ 
belonging to a $m$-fold degenerate
eigenvalue $0$.

The degeneracy also follows from the backward equation
which is governed by the adjoint Fokker-Planck operator $L^+$ defined in 
eq.~(\ref{BFP}).
In the deterministic case the diffusion matrix vanishes, so
the second derivatives disappear and the
stationary solutions of the backward operator are constant along the
trajectories of the deterministic system. Because within a domain of
attraction all trajectories approach
the same attractor the stationary solutions of the noiseless
backward equation are constant on  the domain of attraction 
${\cal D}_\a$.  One then can choose the
characteristic functions $\c^{(0)}_\a(\bx)$
of the domains of attraction as $m$ independent
solutions of the stationary deterministic backward equation 
$K^{(0)}_i(\bx) \partial \c^{(0)}(\bx)/\partial x_i = 0$: 
\be 
\c^{(0)}_\a(\bx) = \left \{ \begin{array}{ll}
1 \qquad &\mbox{for} \; \bx \in {\cal D}_\a  \\
0 & \mbox{else}
\end{array}
\right.
\ee{chi}
In the presence of weak noise these
functions can be modified slightly  
such that they still solve the stationary backward equation up to
corrections that are exponentially small in the noise strength: The
step-like
discontinuities then are  smeared out on  layers located at the
boundaries of the domains of attraction
${\cal D}_\a$. The functions can be constructed by means of the method
of matched 
asymptotic expansions \cite{MS}, see also Refs. \cite{Ry,Mo}. 
Outside the boundary layers 
no modification is required and hence they retain their
property of localizing the different domains of attraction. 
In the following, these localizing functions will be
denoted by $\c_\a(\bx)$. We summarize their main properties: They solve
$L^+\c_\a(\bx) =0$ up to exponentially small corrections in the noise strength,
have the values one within and zero outside ${\cal D}_\a$ and 
interpolate
between these values smoothly on a thin boundary layer \cite{t87,t94}.
As solutions of these particular boundary value problems, the functions
$\c_\a(\bx)$  are the {\it splitting
probabilities} that give the relative frequencies with which a trajectory
starting at $\bx$ first reaches the attractor $\a$ before it visits any other
attractor \cite{Ma}.

The splitting probabilities in general are not eigenfunctions of the backward
operator. However, they span the subspace of eigenfunctions that emerges
from the eigenspace of the 
$m$-fold degenerate eigenvalue $0$ of the deterministic dynamics
if weak noise perturbes the
system and lifts the degeneracy.
The resulting finite eigenvalues are still small compared to all
other finite eigenvalues. Hence, this subspace describes the slow dynamics of
the system at weak noise. In particular, it contains the constant function
which is the exact eigenfunction of the backward equation belonging to the
eigenvalue $0$. At weak noise this eigenfunction
can be represented as the sum over the characteristic
functions $\c_\a(\bx)$:
\be
\sum_\a \c_\a(\bx) = 1.
\ee{s1}

Because of detailed
balance, the slow subspace of the forward operator is spanned by the basis
set $\{ \tilde{\c}_\a(\bx) \r_0(\bx) \}$, where $\tilde{\c}_\a(\bx) =
\c_\a(\tilde{\bx})$ denotes
the image of $\c_a(\bx)$ under time
reversal. Hence, the probability density describing the slow
dynamics can be represented as:
\be
\r(\bx,t) = \sum_{\a =1}^m c_\a(t) \tilde{\c}_\a(\bx) \r_0(\bx)
\ee{sr}

As already noted the characteristic functions $\c_\a(\bx)$ and, therefore also
their time reversed partners $\tilde{\c}_\a(\bx)$, are well localized functions
as long as the noise does not become too large.
As a consequence, one may
use the functions $\c_\a(\bx)$ in order to determine the probability $p_\a(t)$
that the system resides in the domain of attraction of the attractor $\a$
\cite{Mo}:
\be
p_\a(t) = \left ( \c_\a, \r(t) \right )
\ee{pat}
where the scalar product is defined as the integral of the product of its
arguments over the state space:
\be
\left ( f, g \right ) = \i_\S d^n\bx f(\bx) g(\bx).
\ee{sp}
If we assume that the probability density $\r(t)$  is of the form (\ref{sr})
 we obtain
\be
p_\a(t) = c_\a(t) n_\a
\ee{pn}
where we have made use of the localization of the characteristic
functions at weak noise:
\be
\left ( \c_\a,\tilde{\c}_{\a'} \r_0 \right ) = \d_{\a,\a'} n_\a
\ee{cab}
and have introduced the function 
$n_\a = \i dx \c_\a(x)\tilde{\c}_{\a}(x) \r_0(x) 
\approx \i dx \c_\a(x) \r_0(x)$ 
giving the
population of the metastable state $\a$ in the equilibrium distribution 
$\r_0(x)$. 

The time evolution of the probabilities $p_\a(t)$ now follows from the
projection of the Fokker-Planck equation (\ref{FP})
onto the slow subspace, i.e. we
assume the probability density to take the form (\ref{sr}) and determine
the scalar product with $\c_\a(\bx)$ on both
sides of the Fokker-Planck equation.
This
yields
\be
\frac{d}{d t} p_\a(t) = \sum_{\a'}
\frac{ \left ( \c_a,L \tilde{\c}_{\a'} \r_0 \right ) } {n_{\a'}} p_{\a'}(t)
\ee{pme}
Under the sum on the right hand side 
we expressed the coefficients $c_{\a'}(t)$ by means of
eq.~(\ref{pn}) in terms of the
probabilities $p_{\a'}(t)$ leading to the equilibrium weights $n_{\a'}$ in the
denominator of the respective coefficient.
For $\a \neq \a'$ the coefficient $\left ( \c_\a, L \tilde{\c}_{\a'}
\r_0 \right )/n_{\a'} $ coincides with the Rayleigh-quotient
expressions for the rate from $\a'$ to $\a$ \cite{t94}. 
The integral is dominated by a
neighborhood of the saddle
point that connects the domains of attraction $\mathcal{D}_{\a'}$ and 
$\mathcal{D}_\a$ \cite{c1}. Within this neighborhood, for small noise, 
the localizing function $\c_\a(\bx)$ can be
approximated by the characteristic function of $\cal{D}_\a $,
$\c_\a^{0}(x)$ \cite{t95}. 
The resulting coefficient then
coincides with the better known flux-over-population expression for
the respective rate \cite{htb}. In particular, 
one can show that for sufficiently weak noise the expression 
$\left ( \c_\a, L \tilde{\c}_{\a'}\r_0 \right )/n_{\a'} $ is positive
for $\a \neq \a'$.     
From eq.~(\ref{s1}) it
follows that the sum of the coefficients over
$\a$ vanishes. Hence, eq.~(\ref{pme}) has the proper form of a master
equation:
\be
\frac{d}{d t} p_\a(t) = \sum_{\a' \neq \a} r_{\a,\a'} p_{\a'}(t) -
\sum_{\a' \neq \a} r_{\a'\!,\a}\: p_\a(t)
\ee{me}
where
\be
r_{\a,\a' } = \frac{ \left ( \c_\a, L \tilde{\c}_{\a'} \r_0 \right ) }{n_{\a'}}
\ee{rab}
is the transition probability per unit time from $\a'$ to $\a$.

The regime of validity of this approximation, of course, depends on the
particular system and the degree of accuracy one would like to achieve.
By its very nature this master equation 
is only appropriate for the long time dynamics of the
Fokker-Planck equation and therefore requires a clear separation of time
scales dividing the process into fast relaxations within the domains of
attraction and slow transitions between the attractors. We note that various
definitions of transition rates exist 
which are different from a physical point of
view
but which still yield the same result if a master equation provides a
correct description of the long time dynamics. It turns out that the actual
differences between these rate expressions are exponentially small in the
noise strength, i.e. of the form $\exp \left \{ -\D V/k_B T \right \}$ where
$\D V $ is a relevant barrier hight and $k_B T$ the noise strength. So if we
are ready to accept an error of, say $1 \%$, all barriers in the system must
be higher than approximately $4.5 k_B T$.

\section{Slow driving}\label{III}
After these preparatory considerations we turn to our main topic and
consider a Fokker-Planck process of the kind of the previous section with the
only difference that we now allow for slow changes of the parameters of the
system. These could result from a slow drift of the environmental parameters
of the system like e.g. the temperature, a slow increase of an external field,
or a slow periodic force driving the system, to name but a few
realizations.

We emphasize that the change of the parameters must be slow compared to
the fast relaxation within each domain of attraction but no assumption is
made how it compares to the slow intrinsic dynamics of the system describing
the transitions between different domains of attraction.
We assume that in the presence of these time dependent
parameters one still can
identify domains of attraction which do not merge or split as time changes
but retain their identity. In other words, we restrict ourselves to changes
of the parameters that do not lead to topological changes of the system.

The resulting process is still Markovian and is governed by a
Fokker-Planck equation:
\be
\frac{\partial}{\partial t} \r(t) = L(t) \r(t)
\ee{FPT}
where the Fokker-Planck operator $L(t)$ is of the form given in eq.~(\ref{FPO})
with time dependent drift $\bK(\bx,t)$ and 
possibly also a time dependent diffusion
matrix $\bD(\bx,t)$.
In particular, we assume that all parameter values which the system
experiences in the course of time correspond to equilibrium systems, i.e.
that for any frozen parameter value the system will reach an equilibrium
distribution 
relative to which the system obeys detailed balance:
\be
L(t) \hat{\r}_0(t) = \hat{\r}_0(t) \tilde{L}^+(t)
\ee{fdb}
where the tilde denotes time reversal and $\hat{\r}_0(t)$ is the
multiplication operator with the equilibrium distribution
$\r_0(\bx,t)$ 
that would be attained if the parameter values were frozen at the
values that they assume at time $t$:
\be
L(t) \r_0(\bx,t) =0.
\ee{L0}

For vanishing noise the system will move toward the nearest attractor
as defined by the momentary values of the parameters. During this relaxation
time, the
parameters of the system are supposed to change by such a small amount 
that they can be
considered as kept constant. Only when the system has reached the attractor
and stays there for a much longer time,
the change of parameters will become noticeable and the
system will follow the slow
motion of the attractor as it results from the changing parameters.
This means that the time rate of change of the deterministic drift is
characterized by a small parameter $\e$:
\be
\frac{|\dot{\bK}(\bx,t)|}{| \bK(\bx,t)|} \t_r= {\cal O}(\e)
\ee{e}
where $\t_r$ is a typical relaxation time of the deterministic motion.
Under this very condition many of the properties of the time homogeneous
process carry over to the case with time dependent parameters.

To each attractor $\a$ then there belongs a domain of attraction
${\cal D}_\a(t)$ together with its characteristic function 
$\c^{(0)}_\a(\bx,t)$ which is
$1$ on ${\cal D}_\a(t)$ and $0$ elsewhere in $\S$. The domains of attractions
also slowly depend on time and so do the characteristic functions.
If the system
is weakly perturbed by noise the characteristic functions will change into
smooth functions $\c_\a(\bx,t)$
that coincide with the noiseless functions everywhere
except in a thin boundary layer at $\partial {\cal D}_\a(t)$, where a steep
but smooth transition from $1$ to $0$ takes place. As in the time
independent case, they are solutions of the homogeneous backward equation that
approach $1$ in the interior of ${\cal D}_\a(t)$ and vanish on all other
attractors $\a' \neq \a$.
Further one also can represent the slow
dynamics of the probability
density in terms of linear
combinations of ${\tilde \c}_\a(\bx,t) \r_0(\bx,t)$:
\be
\r(\bx,t) = \sum_{\a =1}^m c_\a(t) \tilde{\c}_\a(\bx,t) \r_0(\bx,t)
\ee{srt}
The probabilities to find the system in the state $\a$ is
analogously defined as in the case with constant parameters:
\be
p_\a(t) = \left (\c_\a(t), \r(t) \right )
\ee{patt}

The time rate of change of $p_\a(t)$ consequently has two contributions,
resulting from the derivative of $\c_\a(\bx,t)$ and of $\r_0(\bx,t)$:
\be
\frac{d}{dt} p_\a(t)= \left ( \frac{\partial}{\partial} \c_\a(t), \r(t) \right ) + 
\left ( \c_\a(t), \frac{\partial}{\partial t}\r(t) \right ).
\ee{dpa}
The latter probabilistic contribution
can be expressed in terms of the Fokker-Planck equation yielding
\be
\left ( \c_{\a}, \frac{ \partial }{ \partial t} \r(t) \right )
= \sum_{\a'} r_{\a,\a' }(t) p_\a (t)
\ee{pt1}
where we used eq.~(\ref{srt}) and introduced the time dependent
transition 
rates $r_{\a,\a' }(t)$ in an analogous way as in the time homogeneous
case, see eq.~(\ref{rab}):
\be
r_{\a,\a' }(t) = \frac{ \left ( \c_\a(t), L(t) 
\tilde{\c}_{\a'} (t) \r_0(t) \right )}{
n_{\a'} (t) }.
\ee{rt}
It determines the rate of change of the probability $p_\a(t)$ caused by
transitions from $\a'$ to $\a$. Here $n_{\a'}(t)$ is the population of the
metastable state $\a'$ in the frozen equilibrium distribution $\r_0(\bx,t)$. It
is given by:
\be
n_\a(t) = \left ( \c_\a(t), \tilde{\c}_\a(t) \r_0(t) \right ) \approx
 \left ( \c_\a(t), \r_0(t) \right )
\ee{nat}
where the term $\tilde{\c}_\a(\bx,t)$ is neglected in the second
equation. This approximation 
holds up to exponentially small terms in the noise
strength.

The other contribution, $\left (\partial \c_\a (t) / \partial t,\r(t) \right
)$,
describes the change of the probability caused by the geometric
change of the domain ${\cal D}_\a(t)$. It also is linear in the probabilities
$p_{\a'} (t)$ and can be written as
\be
\left (\frac{\partial}{ \partial t} \c_\a(t), \r(t) \right ) = \sum_{\a'}
g_{\a,\a' }(t) p_{\a'} (t)
\ee{pt2}
where
\be
g_{\a,\a' }(t) = \frac{ \left ( \frac{\partial}{ \partial t} \c_\a(t),
\tilde{\c}_{\a'} (t) \r_0(t) \right )}{ n_{\a'} (t) }
\ee{g}
The coefficients $g_{\a,\a' }(t)$ are proportional to the smallness parameter
$\e$ because they contain the time derivative of $\c_\a(x,t)$. The proportionality
factor roughly is of the same order of magnitude as the transition rate
$r_{\a,\a'}(t)$: 
\be
\frac{ g_{\a,\a' }(t)}{ r_{\a,\a' }(t)} = {\cal O}(\e)
\ee{gr}
The dependence of this ratio on the noise strength is discussed for a
particular example in section~\ref{V}. 
For a sufficiently slow parameter change, i.e a small value of $\e$, 
the geometric contribution to the time rate of change of the
probability $p_\a(t)$ can be neglected and a master equation results that
is completely determined by the instantaneous rates $r_{\a,\a'}(t)$:
\be
\frac{d}{d t} p_\a(t) = 
\sum_{\a'\neq \a} r_{\a,\a'}(t) p_{\a'}(t)-\sum_{\a'\neq \a}
r_{\a',\a}(t) 
p_{\a}(t)
\ee{met}
where we used that the time dependent rates $r_{\a,\a'}(t)$ (\ref{rt}) 
have the same 
general form and therefore   
the same formal properties
as the time independent ones (\ref{rab}): 
Those for $\a \neq \a'$ are positive and the
sum over the first index vanishes: $\sum_\a r_{\a,\a'}(t)= 0$.
This master equation with time dependent rates is the central result
of the present paper. 
For faster driving the geometric contributions of the rates have to be taken 
into account. However, they may become negative, and therefore formally 
negative rates may result if the driving is too fast. This indicates that then 
the instananeous eigenfunctions of the Fokker Planck equation no longer 
provide a good basis.

\section{Decorating the meta-stable states}\label{IV}
Often the knowledge of the probabilities $p_\a(t)$ to find the system at time
$t$ in the meta-stable state $\a$ is not sufficient. For example one
may be interested in the average position $\bx$ of the system or its
single- or multi-time statistical properties.
We here show how the time dependence of these quantities on the slow time
scales of the transitions
and the external driving can be retrieved from exactly the same information
that is necessary to determine the transition rates from the Fokker-Planck
equation. For that purpose we first
consider the average of an arbitrary function
$f(\bx)$
of the position:
\ba
\langle f(t) \rangle & = & \i d^n \bx f(\bx) \r(\bx,t) \nonumber \\
& = & \sum_\a \langle f(t) |\a \rangle p_\a(t)
\ea{fa}
where $\langle f(t) |\a \rangle$ denotes the expectation value of $f(x)$ under
the condition that the system resides at $t$ in the domain of attraction ${\cal
D}_\a(t)$. Using the long-time behavior of the probability density $\r(t)$ as
given by eq.~(\ref{srt}) one finds:
\be
\langle f(t) |\a \rangle  = \i d^n \bx f(\bx) \r(\bx,t|\a)
\ee{rxta}
where the conditional probability $\r(\bx,t|\a)$ to find the system in the
continuous state $ \bx \in {\cal D}_\a(t)$  is given by
\be
\r(\bx,t|\a) =  \frac{\tilde{\c}_\a(\bx,t)}{ n_\a(t)} \r_0(\bx,t)
\ee{rxtae}
Similarly, one finds for the correlation of two functions $f(\bx)$ and $g(\bx)$
at times $t$ and $t +\t$ being separated by a positive time $\t$ that is long
compared to the
fast time scale of relaxations within the domains of attraction the
following expression:
\be
\langle f(t+ \t) g(t) \rangle = \sum_{\a,\a' } \langle f(t+\t)|\a \rangle\:
\langle g(t) |\a' \rangle\: p(\a, t+\t | \a', t )\: p_{\a'}(t).
\ee{fg}
Here $p(\a,t+\t|\a',t)$ is the conditional probability to find the
system in the metastable state $\a$ at time $t+\t$ provided it was in the
state $\a'$ at the earlier time time $t$. This conditional probability is the
solution of the master equation (\ref{met}) subject to the initial
conditions $p(\a, t| \a', t) = \d_{\a,\a'}$. Eq.~(\ref{fg}) holds for 
arbitrary functions $f(\bx)$ and $g(\bx)$. Therefore
the conditional probability of 
finding the system at time $t+\t$ at the continuous state $x$  if it was at 
the earlier time $t$ at $y$ takes the form
\be
\r(\bx,t+\t|\by,t) = \sum_{\a,\a'} \r(\bx,t+\t|\a) \:\c_{\a'}(\by,t) \: 
p(\a,t+\t|\a',t).
\ee{cpx}
For an independent derivation of this result see the Appendix~\ref{B}.

Yet another analogous way exists 
to characterize the continuous process on the long time scale: 
One takes the discrete process $z(t)$ assuming the values 
$\a = 1\ldots m$ 
according to the master equation (\ref{met}) and decorates the 
states with a random point in the state space, 
$\bX(z(t),t)$, depending on time and the 
particular state  $\a=z(t)$ that is realized at the time $t$.
If the probability density of the decoration $\bX(\a,t)$ is chosen as the
conditional
probability density $\r(\bx,t|\a)$ given in eq.~(\ref{rxtae}) the mean values
and the correlation functions of the process
$\bX(\bx,t)$ 
coincide with the expressions (\ref{fa})
and (\ref{fg}), respectively, and accordingly are characterized by the
conditional probability (\ref{cpx}). The decorated process is
Markovian and has the same single and two time probability density as
the continuous process $\bx(t)$ on the long time scale and hence
coincides there 
with it: 
\be
\bx(t)= \bX(z(t),t) 
\ee{xXz}

\section{Bistable overdamped oscillator}\label{V}
We here consider the example of an overdamped bistable Brownian oscillator
that is periodically driven at a frequency $\O$ which is slow compared to the
typical relaxation rates. Its dynamics is given by the Smoluchowski operator
\be
L_S(t) = \frac{\partial}{ \partial x} \frac{\partial V(x,t)}{ \partial x} +
\b^{-1} \frac{\partial^2}{ \partial x^2}
\ee{LS}
where $\b^{-1}$ is the noise strength which is proportional to the
temperature $\th$ of the fluid surrounding the oscillator, 
and  $V(x,t)$ denotes the time-dependent potential:
\be
V(x,t) = \frac{1 }{ 4} x^4 - \frac{1 }{ 2} x^2 - A x \sin(\O t).
\ee{Vxt}
Here $A$ denotes the strength of the periodic force. Throughout dimensionless 
units are used.
It is supposed to stay within the limits $|A| < 2/(3 \sqrt{3) }$. Actually
it must keep some finite distance from these limits in order that the
following asymptotic theory applies. Under this condition on the strength of
the external force, the potential
has three stationary points that are solutions of the algebraic equation:
\be
x^3 -x = A \sin(\O t)
\ee{V'}
As time varies they form three branches: $x_1(t)$ and $x_2(t)$ are those
tracing the two local minima that at $t=0$ assume the values $x_1(0) = -1$
and $x_2(0) = 1$, respectively, and $x_b(t)$ gives the location of the local
potential maximum between the minima. At $t=0$ it is located at $x_b(0) =0$.
The corresponding extreme values of the potential $V(x,t)$
are denoted by:
\ba
V_\a(t) & = & V(x_\a(t),t) \qquad \mbox{for} \;\a=1,2 \nonumber \\
V_b(t) & = & V(x_b(t),t)
\ea{Vpmb}
The instantaneous well and barrier frequencies are defined accordingly as:
\ba
\o_\a(t) = & \sqrt{\frac{ \partial^2 V(x_\a(t),t)}{ \partial x}} = & \sqrt{3
x_\a^2(t) -1} \qquad \mbox{for} \; \a=1,2\nonumber \\
\o_b(t) = & \sqrt{\frac{ -\partial^2 V(x_b(t),t)}{ \partial x}} = & \sqrt{1-3
x_b^2(t)}
\ea{opmb}
For later use we give the time derivatives of the barrier position and the
barrier 
frequencies. For the position one finds from eq.~(\ref{V'}):
\be
\dot{x}_b (t)  =  -\frac{ \O A}{ \o_b(t)^2} \cos(\O t)
\ee{dx}
and for the frequency it follows:
\be
\dot{\o}_b(t) = -3\frac{ x_b(t)}{ \o_b(t)} \dot{x}_b(t)
\ee{do}
The instantaneous stationary solution of the Smoluchowski equation 
$L_S(t) \r_0(x,t) = 0$ is given by the Boltzmann distribution:
\be
\r_0(x,t) = Z^{-1}(t) e^{-\b V(x,t)}
\ee{rB}
where for small noise the partition function is given by
\be
Z(t) = \sqrt{\frac{2 \p }{ \b}} \left \{ 
\frac{ \exp \left \{-\b V_1(t)\right \} }{
 \o_1} +
\frac{\exp \left \{-\b V_2(t)\right \} 
}{ \o_2} \right \}
\ee{Z}
Here algebraic correction terms of the order ${\cal O}(1/(\b \Delta
V_\a(t)))$ were neglected.
The domains of attraction of the instantaneous locally stable states
$x_\a(t)$, $\a = 1,2$, 
extend from minus infinity to the instantaneous barrier and
from there to infinity: ${\cal D}_1(t) = ( -\infty,x_b(t))$, and
${\cal D}_2(t) = ( x_b(t), \infty) $.

The corresponding localizing functions $\c_\a(x,t)$
are the solutions of the backward equation
\be
\left \{\o_b(t)^2 (x-x_b(t)) \frac{\partial}{ \partial x} + \b^{-1}
\frac{\partial^2}{ \partial x^2}  \right \} \c_\a(x,t) = 0
\ee{ca}
with boundary conditions:
\be \begin{array}{c}
\c_1(x,t)  =  \left \{\begin{array}{ll}
1 \quad &\mbox{for}\;\; x \to - \infty \\
0 &\mbox{for}\; \; x \to \infty
\end{array} \right .\\
 \\
\c_2(x,t)  =  \left \{\begin{array}{ll}
0 \quad &\mbox{for}\;\; x \to - \infty \\
1 &\mbox{for}\; \; x \to \infty
\end{array} \right .
\end{array}
\ee{c12}

The solutions readily are found:
\ba
\c_1(x,t)& = & \frac{1 }{ 2} \mbox{erfc}(\o_b(t)\sqrt{\b/2}(x-x_b(t)))
\nonumber \\
\c_2(x,t)& = & 1 - \c_1(x,t)
\ea{cc}

In the present model there is only a single variable, the coordinate
$x$ 
which transforms evenly under time reversal, hence, $\tilde{\c}_\a(x,t)
= \c_\a(x,t)$.
Using the eq.~(\ref{cc}), the scalar products
$\left ( \c_\a(t), L_S(t) \c_{\a'}(t) \r_0(t) \right )$, $\a,\a' = 1,2$ can be
expressed by a single one that we denote by $q(t)$:
\ba
\left (\c_1(t), L_S(t) \c_2(t) \r_0(t) \right )& = &
\left (\c_2(t), L_S(t) \c_1(t) \r_0(t) \right ) =  \nonumber \\
-\left (\c_1(t), L_S(t) \c_1(t) \r_0(t) \right )& = &
-\left (\c_2(t), L_S(t) \c_2(t) \r_0(t) \right ) =  q(t)
\ea{cq}
where $q(t)$ takes the form:
\be
q(t) = \b^{-1} \i dx \left ( \frac{\partial \c_1(x,t)}{ \partial x}
\right )^2 \r_0(x,t)
\ee{qt}
Using eq.~(\ref{cc}) for weak noise it can further be simplified as
\be
q(t) = \frac{\o_b(t) e^{-\b V_b(t)}}{ Z(t) \sqrt{2 \p \b}}
\ee{q}

The populations $n_\a(t)$ of the metastable states
$\a=1,2$ in the frozen equilibrium distribution become at weak noise:
\be
n_\a(t) = \sqrt{\frac{2 \p }{ \b}} \frac{ e^{-V_\a(t) \b}}{ \o_\a(t) Z(t)}
\ee{n12t}
Using eq.~(\ref{rt}) this gives for the rates the instantaneous
expressions \cite{htb}:
\ba
r_{2,1}(t) = & - r_{1,1}(t) = & \frac{\o_1(t) \o_b(t)}{ 2 \p} e^{-\b \D V_1(t)}
 \nonumber \\
r_{1,2}(t) = & - r_{2,2}(t) = & \frac{\o_2(t) \o_b(t)}{ 2 \p} e^{-\b \D V_2(t)}
\ea{r12t}
\subsection{The geometric correction to the rate}\label{Va} 
\begin{figure}
\begin{center}
\includegraphics[width=8cm]{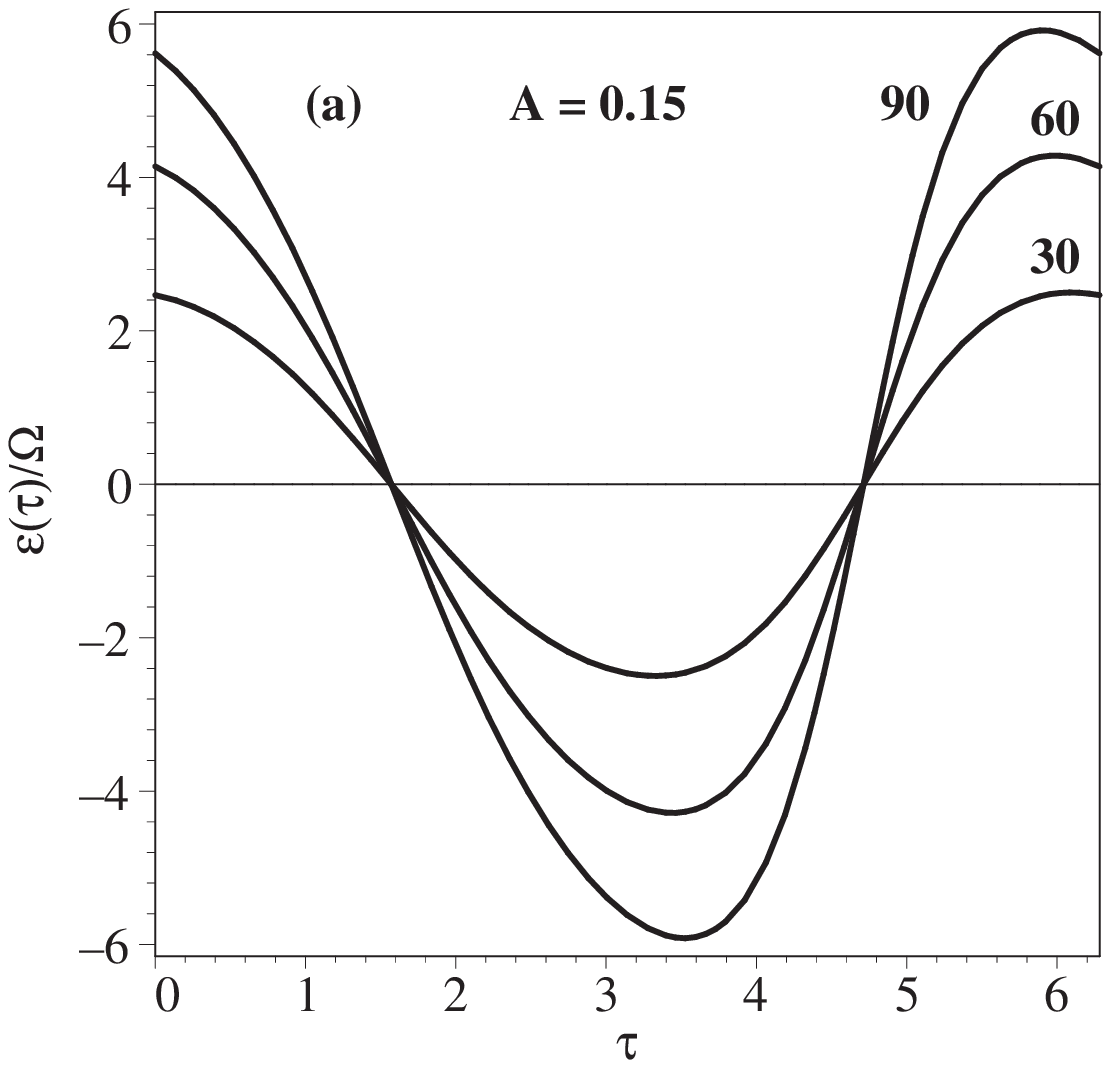}
\hfill
\includegraphics[width=8cm]{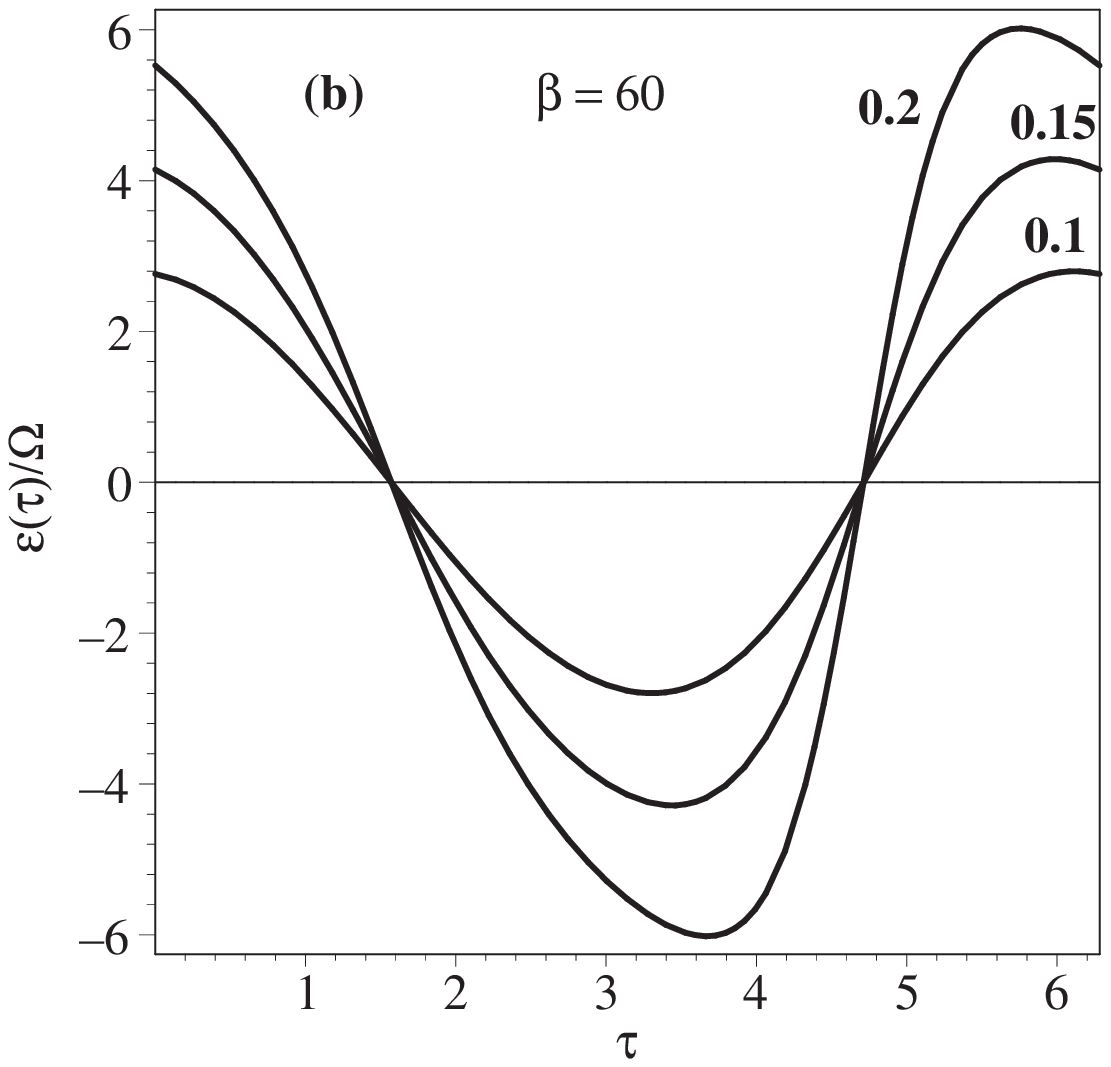}
\end{center}
\caption{The relative error of the rate oscillates as a function of
  time $\t = \O t$. It vanishes at the extrema of the driving force. Panel
  (a) shows $\e(\t)/\O$ for different inverse temperatures $\b = 30,\:
  60, \: 90$ at the
  driving strength $A = 0.15$ and panel (b) for different driving
  strength $A = 0.1,\: 0.15,\: 0.2$ at $\b = 60$. The parameter 
  values are indicated close to the respective curves.}
\label{f2}
\end{figure}

Now we come to the discussion of the corrections of the instantaneous rates 
that are determined by the
geometric contributions $\left ( \partial \c_\a(t)/\partial t,
\c_{\a'}(t) \r_0(t) \right )$, $\a, \a' = 1,2$. 
For the relative magnitude of the geometric correction to the
instantaneous rate $r_{2,1}(t)$ one finds after some algebra:
\ba
\e(t)& \equiv & \frac{g_{2,1}(t)}{r_{2,1}(t)}  = 
\frac{\left ( \partial \c_2(t)/ \partial t,\c_1(t) \r_0(t) \right
  )}{\left (\c_2(t), L_S(t) \c_1(t) \r_0(t)\right )} \nonumber \\ 
& = & \ - \frac{\b}{2} 
\dot{x}_b(t) 
\i_{-\infty}^\infty dy \left (\frac{3 x_b(t)}{\o_b^2(t)} y +1 \right ) 
\nonumber \\ & &\times 
\mbox{erfc} \left ( \o_b(t) \sqrt{\b/2} y \right ) 
\exp \left \{ -\b \left ( y^4/4 +
      x_b(t) y^3 \right ) \right \} 
\ea{g21}
As expected, the error is proportional to the driving frequency
$\O$ via the time derivative $\dot{x}_b(t)$. 
The remaining $\O$-dependence may be absorbed in a rescaled time
$\t= \O t$.  
An analytic solution of the integral in eq.~(\ref{g21}) is not known. 
In Fig.~\ref{f2}
the error divided by the driving frequency 
is shown as it results from a numerical evaluation for different values
of the driving strength $A$ and inverse temperature as a function of
time. Both increasing driving strength $A$ and inverse temperature
$\b$ lead to an increase of the extrema of the error that must be
compensated by a smaller driving frequency in order that the master
equation with the instantaneous rates provides a valid description. 

\begin{figure}
\includegraphics[width=8cm]{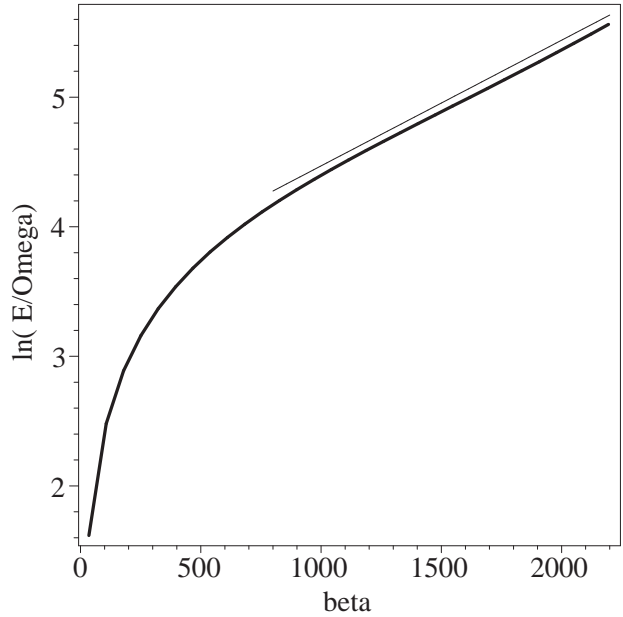}
\caption{The asymptotic exponential growth of the frequency 
independent ratio $E/\O$ is shown for the driving strength
$A=0.12$. The straight line corresponds to a barrier height $0.97 \times
10^{-3}$.}
\label{f3}
\end{figure}

As an average measure we introduce the  root mean square error 
$E= \left ( \i_0^T dt \e^2(t)/ T \right )^{1/2}$ which is  strictly
proportional to the driving frequency $\O$. 
The proportionality factor $E/\O$ increases as a function of the
inverse temperature first as a power law and changes to an exponential
growth for large values of $\b$, see Fig~\ref{f3}. 

We propose the
following physical
explanation of this effect: At large $\beta$ the noise is very small
such that a particle that has reached at some time the top of the
barrier has a very small chance to be pushed to either side by the
noise. During the same time when it resides there, the potential moves
such that the particle no longer will sit on top of the
barrier. Depending on the direction of the motion of the potential the
particle may now be on the opposite side of the barrier or on the
original side where it came from. In the former case the geometric
correction leads to an increase of the rate and in the latter case to
a decrease of the transition rate. Because it then has again to
overcome a barrier the increase of the rate is exponentially large in
the inverse temperature.

Finally we note that the deviation from the frozen
rate on the noise strength depends on the particular model. 
For example in symmetric systems 
with varying barrier height the deviation only grows with the square
root of 
the inverse temperature rather than exponential as in the above case
where both the location and the height of the barrier changes in time.

\subsection{Average motion}\label{Vb}
We have projected the dynamics of a
a  process with multiple metastable states
and slowly time dependent parameters onto the slow subspace defined by
the eigenfunctions of Fokker-Planck operator   

\begin{figure}[b]
\includegraphics[width=10cm]{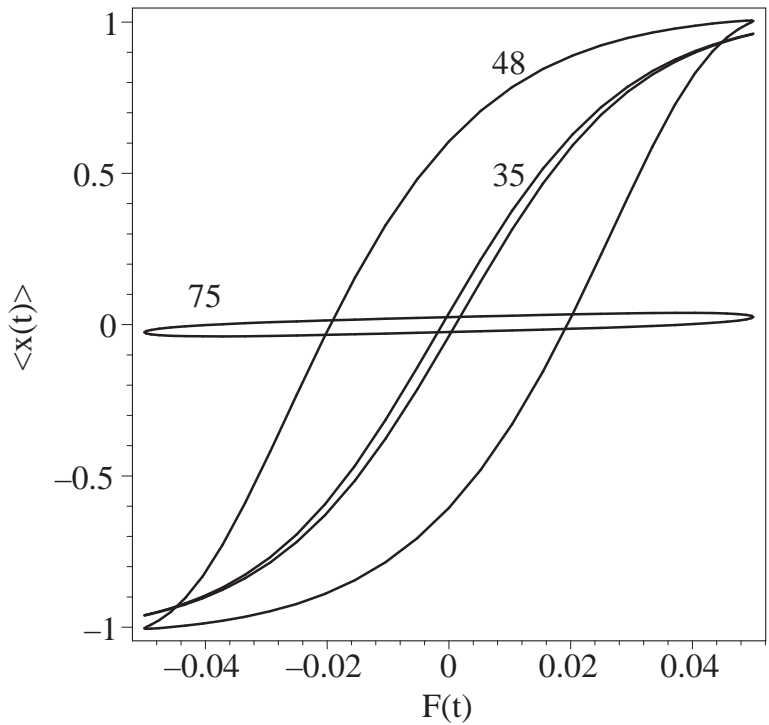}
\caption{The response of the average position on the driving force for
  the driving strength $A=0.05$ and driving frequency $\O = 10^{-5}$
  strongly depends on the magnitude of the noise. For $\beta = 35$ the
  noise is relatively large such that the average position closely
  follows its adiabatic value and therefore shows almost no
  hysteresis; 
  for $\beta = 48$ it stays behind the
  force, and for $\beta = 75$ it is hardly influenced by the force. The
  hysteresis curves are traced in the mathematically positive sense.}
\label{f4}
\end{figure}   
As a simple example of the decoration of the two metastable states we
come back to the bistable oscillator and 
consider the asymptotic motion of its average position. From
eq.~(\ref{fa}) it results as
\be
\langle x(t) \rangle = \langle x(t)|1 \rangle p_1(t) +
\langle x(t)|2 \rangle p_2(t)
\ee{xt}
where $\langle x(t) | \a \rangle $ denotes the conditional average of
the position in the well corresponding to the metastable state $\a$:
\ba
\langle x(t)|\a \rangle & = & \i dx \:x \frac{\c_\a(x,t)}{n_\a(t)} 
\r_0(x,t) \nonumber \\
& \approx & x_\a(t)
\ea{xta}
In the last equation we disregarded small contributions of the order 
$\mathcal{O}(1/(\o_\a(t)\sqrt{\b}))$. The probabilities $p_\a(t)$
follow as the asymptotic solutions of the master equation:
\ba
p_1(t)& = & \frac{\i_0^T ds \:e^{(K(T)-K(s))} r_{1,2}(t)}{1- e^{K(T)}}\: 
e^{-K(t)} +
    \i_0^t ds \:e^{-(K(t) -K(s) )} r_{1,2}(s) \nonumber \\
p_2(t) & = & 1 - p_2(t)
\ea{p1}
where 
\be
K(t) = \i_0^t ds \left ( r_{1,2}(t) + r_{2,1}(t) \right )
\ee{K}

Rather than considering the time dependence we here show in
Fig.~\ref{f4} the dependence of
the position on the driving force $F(t) = A \cos (\O t)$. In general
the position lacks behind the force and therefore the dependence is
hysteretic. Only if the driving is either very slow or very fast 
compared to the average rates there is no hysteretic behavior. 
In the slow case many transitions 
occur before a change of the potential becomes sensible. The
asymptotic probabilities $p_\a(t)$ take the form of the stationary
probabilities for the frozen rates. 
They then depend on the force but not on its rate of change.
In the
opposite limit the asymptotic probabilities see only the average rates
and hence are time independent.        

\section{Summary}\label{VI}
In this paper we studied the dynamics of externally driven systems 
with multiple metastable states in the semiadiabatic limit. The
full dynamics of the system comprising the fast and the slow 
time scales of the intrawell 
relaxations and of the external driving as well as of
the transitions between the metastable states, respectively,
is supposed to be Markovian and continuous and, hence, 
described by a Fokker-Planck equation. We showed that the long
time dynamics of this Fokker-Planck equation is equivalent to the
dynamics of the master
equation for the transitions between the discrete metastable states. 
The rates that determine this master equation are time dependent
taking values as if the parameters were frozen at their instantaneous
values. This kind of kinetic description is not new and 
has long been used in the literature \cite{McNW,ki}. The present work,
however, provides three new aspects. First the master equation does
not result from an educated guess. 
It
rather is 
obtained as the result of the projection of the dynamics onto the slow
subspace of the Fokker-Planck equation. This systematic approach
additionally gives a quantitative criterion when deviations from
the master equation with frozen rates must be expected. This criterion 
does explicitly take into account the time change of the parameters 
and in contrast to other works \cite{Ast} is not
based on a comparison of time scales of the
frozen dynamics with the rate of change of the parameters. As a
particular result we found for the dynamics of a periodically driven 
overdamped Brownian oscillator that the maximal driving frequency
depends on the noise strength and decreases with decreasing noise
strength. Asymptotically this dependence even is exponential. Future work
still has to show whether one can achieve an improvement of the
dynamics if the so called geometric corrections to the rates are taken
into account in
cases when the time scale separation between the fast intrawell
relaxation and the changes of the parameters becomes less pronounced.     
The third main outcome of the present approach is a partial retrieval
of the underlying dynamics of the continuous variables by means of a
proper decoration of the discrete states with time dependent random
variables.

The necessary time scale separation forced us to
exclude the occurrence of bifurcations of the deterministic dynamics
caused by the change of the parameters. Such bifurcations are
accompanied by a slowing down of the deterministic dynamics
and at the same time by a lowering of a barrier height and
consequently by a decreasing time scale of transitions.    
Hence, for time dependent parameters, 
the time scale separation between interwell and intrawell
dynamics is violated within a time window around the instant of bifurcation.
It seems plausible that it should be possible to cut out this time
window and to bridge it by some connection condition for the
probabilities of those states that are involved in the bifurcation.

Apart from the separation of the time scales, another 
assumption was made in the present work. First, we assumed that       
for fixed values of the parameters the system reaches a state of
thermal equilibrium and hence it obeys detailed balance. We think
that this assumption is not really essential for our results but it
simplifies the analysis considerably. Without detailed balance more
general limit sets of the dynamics can occur apart from limit points such as
limit cycles and chaotic states. In general, the stationary
densities are not known in absence of detailed balance, the
weak noise asymptotics is plagued by notorious nonanalyticities 
\cite{gra}, and rather little is known about transition rates in
nonequilibrium systems. On the other hand there are many important 
systems that are driven out of equilibrium by time dependent
parameters.    

\acknowledgments
The authors thank Igor Goychuk, Peter H\"anggi, Sigmund Kohler, 
Marcin Kostur and Michael Schindler
for valuable discussions and hints. This work was supported by the
Deutsche Forschungsgemeinschaft (SFB438).

\appendix
\section{The conditional probability at long times}\label{B}
For a sufficiently long time lag $\t$ the conditional 
probability $\r(x,t+\t|y,t)$ of the 
continuous process $x(t)$ can be expressed in terms of the basis functions 
$\c_\a(y,t)$ and $\tilde{\c}_\a(x,t+\t) \r_0(x,t +\t)$ 
spanning the slow subspaces of the 
backward operator at time $t$ and the forward operator at time $t+\t$,
respectively. One therefore can write:
\be
\r(x,t+\t|y,t) = \sum_{\b,\b'} d(\b,\b';t,\t) \r(x,t+\t|\b) \c_{\b'}(y,t)
\ee{qrc}
where we expressed the basis function $\tilde{\c}_\a(x,t+\t) \r_0(x,t +\t)$
by $\r(x,t+\t|\a) n_\a(t+\t)$, see eq.~(\ref{rxtae}) and introduced the yet 
undetermined
coefficients $d(\a,\a';t,\t)$ into which the time dependent factor $n_\a(t)$
is absorbed. We now multiply the conditional probability of
the continuous process $\r(x,t+\t|y,t)$ with
$\r(y,t|\a')$, integrate over the domain of attraction ${\cal D}_\a(t+\t)$
and obtain the conditional probability of the discrete process
$p(\a,t+\t|\a',t)$. Using the ansatz (\ref{qrc}) for $\r(x,t+\t|t)$
we then find for the coefficients
$d(\a,\a';t,\t) = p(\a,t+\t|,\a',t)$. Together with eq.~(\ref{qrc}) this
yields the expression eq.~(\ref{cpx}) for the conditional probability
of the continuous
process as claimed in Sect.~\ref{IV}


\begin{thebibliography}{99}
\bibitem{c1} If there is more than one isolated saddle point on the common
  separatrix of the domains of attraction $\mathcal{D}_{\a'}$ and 
$\mathcal{D}_\a$ the contribution of each saddle point to the
  matrixelement in eq.~(\ref{pme}) can be
  determined separately. More complicated cases need an extra
  treatment, see Refs. \cite{htb,btez}.
\end{thebibliography}
\end{document}